\documentclass[preprint,number,12pt]{elsarticle}

\usepackage{amssymb}
\journal{Theory of computing systems}

\begin{document}
\newtheorem{thm}{Theorem}
\newtheorem{lem}[thm]{Lemma}
\newtheorem{cor}{Corollary }
\newdefinition{defn}{Definition}
\begin{frontmatter}

%% Title, authors and addresses

%% use the tnoteref command within \title for footnotes;
%% use the tnotetext command for the associated footnote;
%% use the fnref command within \author or \address for footnotes;
%% use the fntext command for the associated footnote;
%% use the corref command within \author for corresponding author footnotes;
%% use the cortext command for the associated footnote;
%% use the ead command for the email address,
%% and the form \ead[url] for the home page:
%%
%% \title{Title\tnoteref{label1}}
%% \tnotetext[label1]{}
%% \author{Name\corref{cor1}\fnref{label2}}
%% \ead{email address}
%% \ead[url]{home page}
%% \fntext[label2]{}
%% \cortext[cor1]{}
%% \address{Address\fnref{label3}}
%% \fntext[label3]{}

\title{Extra connectivity measures of 3-ary $n$-cubes\tnoteref{support}}
\tnotetext[support]{This work is supported by the Postdoc Research Fund of China and the National Natural Science Foundation of China under Grant Nos.60603098\  \& \ 60804021\  \& \ 60703118 }
%% use optional labels to link authors explicitly to addresses:
%% \author[label1,label2]{<author name>}
%% \address[label1]{<address>}
%% \address[label2]{<address>}

\author[xidian]{Qiang Zhu\corref{cor1}}
\ead{qiangzhu@ustc.edu}
\author[xidian]{Xin-Ke Wang}
\ead{wxk1383@126.com}
\author[lgd]{Juanjuan Ren}
\ead{renjj8011@163.com}

\cortext[cor1]{Corresponding author}
%\cortext[cor2]{Principal corresponding author}

%\fntext[fn1]{This is the specimen author footnote.}
%\fntext[fn2]{Another author footnote, but a little more longer.}
%\fntext[fn3]{Yet another author footnote. Indeed, you can have
%any number of author footnotes.}
\address[xidian]{Department of Mathematics, Xidian University, Xi'an, Shanxi 710071, China}
\address[lgd]{School of Management Science, Xian University of Technology, Xi'an Shanxi 710023, China}

%\author{Qiang Zhu \corref{cor1}} \author{Xinke Wang}
%\ead{qiangzhu@ustc.edu}
%\fntext[label2]{xidian university}
%\cortext[cor1]{corresponding author}
%\address{Department of Mathematics, Xidian University, Xi'an, Shaanxi 710071, China \fnref{label3}}
%\fntext[label 3 ]{label3}

\begin{abstract}
%% Text of abstract
The $h$-extra connectivity is an important parameter to measure the reliability and fault tolerance ability of large interconnection networks. The $k$-ary $n$-cube is an important interconnection network of parallel computing systems. The $1$-restricted connectivity of $k$-ary $n$-cubes has been obtained by Chen et al. for $k \geq 4$ in \cite{Chen2007-p1848-1855}. Nevertheless, the $h$-extra connectivity of 3-ary $n$-cubes has not been obtained yet. In this paper we prove that  the 1-extra connectivity  of 3-ary $n$-cube is $4n-3$ for $n\ge 2$ and the 2-extra connectivity  of 3-ary $n$-cube is $6n-7$ for $n\ge 3$.
\end{abstract}

\begin{keyword}
%% keywords here, in the form: keyword \sep keyword
$3$-ary $n$-cubes \sep extra connectivity \sep super connectivity \sep interconnection networks
%% PACS codes here, in the form: \PACS code \sep code

%% MSC codes here, in the form: \MSC code \sep code
%% or \MSC[2008] code \sep code (2000 is the default)

\end{keyword}

\end{frontmatter}

% \linenumbers

%% main text
\section{Introduction}
Concurrent systems  incorporating a large number of processors
have got much development in recent years. Communication network is the critical component of a concurrent supercomputer. Many algorithms are communication rather than processing limited. In some cases,  the
reliability of multiprocessor systems is quite important. Connectivity and edge connectivity are traditional measures of the reliability of a communication network. Later on, restricted connectivity, restricted edge connectivity, super connectivity and super edge connectivity are proposed as generalizations of connectivity and edge connectivity to better measure the reliability of communication networks \cite{Boesch1986-p240-246, Esfahanian1989-p1586-1591, Esfahanian1988-p195-199, Wang2004-p199-205, Wang2008-p587-596, Chen2007-p1848-1855}. The $h$-extra connectivity and $h$-extra edge connectivity of interconnection networks are introduced by F\'abrega and Fiol \cite{F`abrega1994-p163-170} in 1994. The 1-extra connectivity and The 1-extra edge connectivity are super connectivity and super edge connectivity respectively.
Thus, the $h$-extra connectivity and the $h$-extra edge connectivity are generalizations of super connectivity and super edge connectivity. So they can be used to better measure the fault tolerance ability of interconnection networks.
Xu et al.\cite{Xu2007-p222-226} and Zhu et al.\cite{Zhu2006-p111-121} studied the extra connectivity and extra edge connectivity measures of some interconnection networks.

The $k$-ary $n$-cube $Q_{n}^{k}$ has been used in the design of several concurrent computers\cite{Esfahanian1989-p1586-1591, Roth1993-p35-35}.
It has many good topological properties, for example, low degree and diameter, efficient distributed routing algorithms, low cost embedding of other topologies and so on. Some properties of the $k$-ary $n$-cube network have been investigated, such as resource placement, routing \cite{Yoshinaga2004-p49-58}, etc. In this paper, the $h$($h$ = 1, 2)-extra connectivity of the 3-ary $n$-cube $Q_{n}^{3}$ will be determined.

The rest of this paper is organized as follows. Definition and preliminaries are given in section II. Section III and Section IV discuss the 1-extra connectivity and the 2-extra connectivity of 3-ary $n$-cubes respectively. Finally,  we present our conclusions in Section V.

\section{Preliminaries} \label{S2:Preliminaties}
For all the terminologies and notations not defined here, we
follow \cite{1097029}. For a graph $G = (V, E)$ and $S\subset V(G)$
or $S\subset G$, we use $N_G(S)$  to denote the set of
neighboring vertices  of $S$ in $G- S$, when it is easy to know
from the context what $G$ denotes, it is usually simplified with
$N(S)$. We use $A_G(S)$ to denote the union of $S$ and $N_G(S)$.
And similarly  $A_G(S)$ can be simplified with $A(S)$.
That is, $N_G(S)=\{v\in V(G)-S \ |\  \exists\  u\in S \mbox{ such that } (u, v)\in E(G) \},  \ \
 A_G(S)=N_G(S)\cup S$. Given two vertex disjoint graphs $G_1, G_2$ with the same order,
let $M$ is a perfect matching between $V(G_1)$ and $V(G_2)$,
the graph $G_1\bigoplus G_2$ is defined as follows: $V(G_1\bigoplus G_2)=V(G_1) \cup V(G_2)$ and $E(G_1\bigoplus G_2)=E(G_1) \cup E(G_2) \cup M$.

In \cite{F`abrega1994-p163-170}, J.F$\grave{a}$brega and M.A. Fiol introduced the extra connectivity of interconnection networks as below.

\begin{defn} \cite{F`abrega1994-p163-170}
An vertex-set $S \subseteq V(G)$ is called a $h$-extra vertex-cut if $G-S$ is disconnected and every component of $G-S$ has more than $h$ vertices. The $h$-extra connectivity of $G$, denoted by $\kappa_h(G)$, is defined as the cardinality of a minimum $h$-extra vertex-cut.
\end{defn}

In particular, the 1-extra vertex-cut is called as the extra vertex-cut and the 1-extra connectivity is called as the extra connectivity. Clearly, $\kappa_0(G) =  \kappa(G)$ for any graph G if G is not a complete graph.

In this paper, Lee weight and Lee distance will be used to explore the properties of $3$-ary $n$-cubes. And node labels will be written as $x_{1}x_{2} \cdots x_{n}$ rather than $n$-tuples $(x_1, x_2, ..., x_n)$.

\begin{defn}Lee weight \cite{Bose1995-p1021-1030}:
Let $x = x_{1}x_{2} \cdots x_{n}$, where $0 \leq x_{i} \leq k$ for all $1 \leq i \leq n$, be an n-digit radix $k$ vector. The Lee weight of $x$ is defined as $W_{L}(x) =  \sum_{i = 1}^{n} |x_{i}|$, where $|x_{i}| = min(x_{i}, k-x_{i})$.
\end{defn}

\begin{defn}Lee distance \cite{Bose1995-p1021-1030}:
For any two an n-digit radix $k$ vectors $x, y$, the Lee distance between them, denoted by $D_{L}(x, y)$, is the Lee weight of their bit-wise difference (mod $k$). That is, $D_L(x, y) = W_L(x-y)$.
\end{defn}

For example, for $k=n=4$, $x = 0123$, $y = 3210$, $W_{L}(x) = 0+1+2+|4-3| = 4$, $D_{L}(x, y) = W_{L}(x-y) = W_{L}(3113) = |4-3|+1+1+|4-3| = 4$.

In a $k$-ary $n$-cube, denoted by $Q_n^k$, where $k$ is referred to as the radix and $n$ as the dimension, each node can be identified by an $n$-digits radix $k$ address $x_{1}x_{2} \cdots x_{n}$, where $0 \leq x_{i} \leq k$ for $1 \leq i \leq n$.
Two nodes with addresses $x = x_{1}x_{2} \cdots x_{n}$ and $y = y_{1}y_{2} \cdots y_{n}$ in a $k$-ary $n$-cube have
an edge if and only if $D_L(x, y) = 1$. Therefore, each node is connected to two neighboring nodes
in each dimension. So the $k$-ary $n$-cube is an $2n$-regular graph.
The $k$-ary 1-cube is the well-known ring, while the $k$-ary 2-cube and $k$-ary 3-cube are best known as the torus; a variation of the mesh with wraparound connections. Note that when $k = 2$ the network collapses to the well-known hypercube topology. Fig. 1 shows some examples of $3$-ary $n$-cubes.

%%%%%�˴�  ��ͼ
\begin{figure}%[http]
\centering
\scalebox{1.0}{\includegraphics{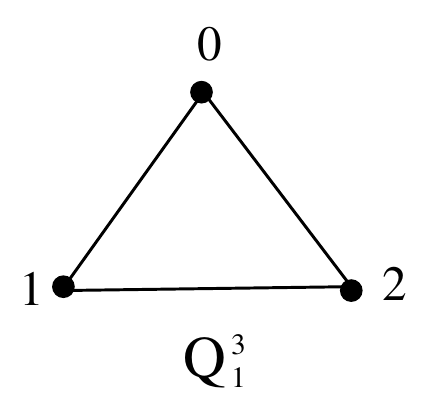}}
\scalebox{1.0}{\includegraphics{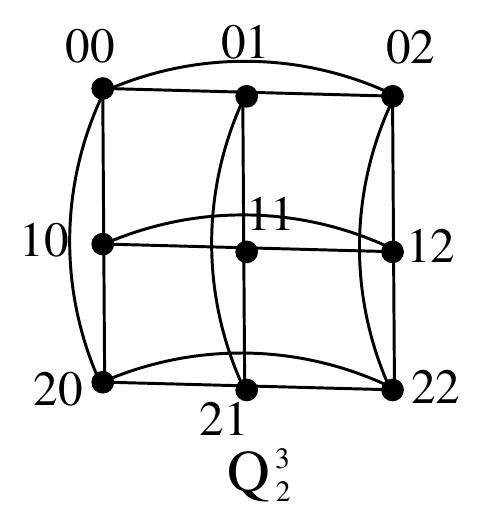}}
\caption{Examples of 3-ary $n$-cubes: $Q_{1}^{3}$ and $Q_2^{3}$}
%\label{fig:anna}
\end{figure}

Given two vertices $x = x_{1}x_{2} \cdots x_{n}$ and $y = y_{1}y_{2} \cdots y_{n}$ in a $k$-ary $n$-cube, the Lee distance $D_L(x, y)$ is also the distance between them. So Lee distance can be used to explore the properties of $k$-ary $n$-cubes.

\begin{lem}\label{dl}
For any two vertices $x, y\in V(Q_n^3)$, $D_{L}(x, y)=l$ iff $x, y$ have exactly $l$ different digits.
\end{lem}
\emph{Proof:}  Suppose $x= x_{1}x_{2} \cdots x_{n}$, $y = y_{1}y_{2} \cdots y_{n}$.
For any $i\in \{1, 2, \cdots, n\}$,
\begin{equation}\label{eq:rayleigh_pdf} %
    y_i -x_i \mbox{\  (mod 3)}=\left\{
        \begin{array}
            {ll}0,&y_i=x_i \\
            1,&y_i\not=x_i
        \end{array}
    \right.
\end{equation}
Since $D_L(x, y) = W_L(x-y)=\sum_{i=1}^n y_i -x_i\mbox{(mod 3)}$,  the result follows. \hfill\rule{1mm}{2mm}

The $i$-th position from the left of the $n$-digit radix 3 string $x_{1}x_{2} \cdots x_{n}$ is called as the $i$-th dimension of $Q_n^3$.
Given an $i$($1\le i\le n$), let $jQ_{n-1}^{3}=\{x_1 x_2\cdots x_n | x_i=j\}$, where $j=0, 1, 2$. Then the spanning subgraph of $jQ_{n-1}^{3}$ is isomorphic to a $3$-ary $(n-1)$-cube. Thus, a $3$-ary $n$-cube can be divided into $3$ vertex-disjoint 3-ary $(n-1)$-cubes, denoted by $Q_n^3=0Q_{n-1}^{3}\bigoplus 1Q_{n-1}^{3}\bigoplus 2Q_{n-1}^{3}$.
Given such a decomposition, by the definition of $Q_n^3$, there is a perfect matching between
$V(iQ_{n-1}^3)$ and $V(jQ_{n-1}^3)$, where $i, j \in \{0, 1, 2\}$
and $i\not=j$.

Given a decomposition of $Q_n^3$, we define the following symbols to simplify some expressions.
For any vertex $u\in V(iQ_{n-1}^{3})$, we call its two neighbors not in  $V(iQ_{n-1}^{3})$ to be $u$'s pair vertices. By the definition of $Q_n^k$, $u$ and its two pair vertices form a triangle in $Q_n^3$.  An edge $(u, v)$ is called an $i$-edge if $u, v$ differs in their $i$-th dimension. In a decomposition of $Q_n^3$ along the $i$-th dimension, all $i$-edges are between different 3-ary $(n-1)$-subcubes, we call these edges as matching edges.

\section{1-extra connectivity}\label{S3:SSG}

\begin{lem}\label{l1}
1) Any two adjacent vertices in $Q_{n}^{3}$ have exactly one common neighbor for $n \geq 1$; \ \ \  2) If any two nonadjacent vertices in $Q_{n}^{3}$ have common neighbors, they have exactly two common neighbors for $n \geq 2$.
\end{lem}
\emph{Proof:}
1) Suppose that $x$ and $y$ are adjacent vertices in $Q_{n}^{3}$. Let $x = x_{1}x_{2} \cdots x_{n}$, then $y = x_{1} \cdots x_{i-1} \bar{x_{i}}x_{i+1} \cdots x_{n}$ with $\bar{x_{i}} = x_{i} \pm 1$ (mod 3). Without loss of generality, assume that $\bar{x_{i}} = x_{i}+1$. Let $z = x_{1} \cdots x_{i-1} \bar{\bar{x_{i}}}x_{i+1} \cdots x_{n}$, where $\bar{\bar{x_{i}}} = x_{i}-1$ (mod 3) $ =  \bar{x_{i}}-2$ (mod 3). Then it is clear that $(x, z) \in E(Q_{n}^{3})$. Since $k = 3$, $D_{L}(y, z) = W_{L}(y, z) = W_{L}(0 \cdots 020 \cdots 0) = |3-2| = 1$. Hence, $(y, z) \in E(Q_{n}^{3})$. Suppose $x$ and $y$ have another common neighbor $u$. Let $u = x_{1} \cdots x_{j-1} \bar{x_{j}}x_{j+1} \cdots x_{n}$ with $j \neq i$. However, $D_{L}(y, u) = W_{L}(y, u) = W_{L}(0 \cdots 010 \cdots 010 \cdots 0) = 2$ and this is a contradiction.

2) Suppose that two nonadjacent vertices $x$ and $y$ in $Q_{n}^{3}$ have a common neighbor $z$. Then the Lee distance between $x$ and $y$ is 2.  Let $x = x_{1}x_{2} \cdots x_{n}$, then $y $ has exactly two digits different from $x$ by Lemma \ref{dl}. We suppose  $y= x_{1} \cdots x_{j-1} \bar{x_{j}}x_{j+1} \cdots x_{i-1} \bar{x_{i}}x_{i+1}  \cdots x_{n}$ with $i \neq j$. Obviously, $x$ and $y$ have exactly two common neighbors $u = x_{1} \cdots x_{j-1} \bar{x_{j}}x_{j+1} \cdots x_{n}$ and $v = x_{1} \cdots x_{i-1} \bar{x_{i}}x_{i+1} \cdots x_{n}$ for $n \geq 2$. $\hfill\rule{1mm}{2mm}$

%\begin{lem}\label{l2}
%Let $S$ be a  vertex subset of $Q_{n}^{3}$ such that $Q_{n}^{3}-S$ is disconnected, then $S_{i} \neq \emptyset$ with $i = 0, 1, 2$ for any decomposition of $Q_n^3$.
%\end{lem}
%\emph{Proof:}
%We use contradiction to prove this lemma. Suppose $S_{0} = {\emptyset}$, where $S_{i} = V(iQ_{n-1}^{3}) \cap S$ with $i = 0, 1, 2$.
%Since there is a perfect matching between $0Q_{n-1}^{3}$ and $1Q_{n-1}^{3}$, for any $u \in V(1Q_{n-1}^{3}-S_{1})$, there is a vertex $v \in V(0Q_{n-1}^{3})$ such that $(u, v) \in E(Q_{n}^{3})$. Therefore, $0Q_{n-1}^{3}\bigoplus 1Q_{n-1}^{3}-S_1$ is connected. By the same argument, $0Q_{n-1}^{3}\bigoplus 2Q_{n-1}^{3}-S_2$ is connected. Hence, $Q_{n}^{3}-S$ is connected. This is a contradiction.
%$\hfill\rule{1mm}{2mm}$

\begin{lem}\label{l3} \cite{Bose1995-p1021-1030}
$\kappa(Q_{n}^{3}) = 2n$.
\end{lem}

\begin{lem}\label{l4}
Let $x$ be an arbitrary vertex in $Q_{n}^{3}$, then $Q_{n}^{3}-A_{Q_{n}^{3}}(x)$ is connected for $n \geq 2$.
\end{lem}
\emph{Proof:}
Given a decomposition of $Q_n^3$: $Q_n^3=0Q_{n-1}^{3}\bigoplus 1Q_{n-1}^{3}\bigoplus 2Q_{n-1}^{3}$. Without loss of generality, suppose $x \in V(0Q_{n-1}^{3})$.  Let $x_1$ and $x_2$ be $x$'s pair vertices in  $V(1Q_{n-1}^{3})$ and $V(2Q_{n-1}^{3})$ respectively. Since $\kappa(Q_{n-1}^{3}) = 2n-2>1$ for $n\ge 2$, both $1Q_{n-1}^{3} - \{ x_1 \}$  and $2Q_{n-1}^{3} - \{ x_2 \}$ are connected.
By the definition of $Q_n^3$,
any vertex in both $V(0Q_{n-1}^{3})-A_{Q_{n}^{3}}(x)$ and $V(2Q_{n-1}^{3}) - \{ x_2 \}$ has a pair vertex in $1Q_{n-1}^{3} - \{ x_1 \}$.
Hence, $Q_{n}^{3}-A_{Q_{n}^{3}}(x)$ is connected. $\hfill\rule{1mm}{2mm}$

\begin{lem}\label{l5}
Let $(u, v)$ be an arbitrary edge in $Q_{n}^{3}$ and $S = N_{Q_{n}^{3}}(u, v)$. Then $|N_{Q_{n}^{3}}(u, v)| = 4n-3$ and $N_{Q_{n}^{3}}(u, v)$ is an 1-extra vertex-cut for $n \geq 2$.
\end{lem}
\emph{Proof:}
It is clear that $Q_{n}^{3}-N_{Q_{n}^{3}}(u, v)$ is disconnected. By Lemma \ref{l1}, $u$ and $v$ have exactly one common neighbor, then $|N_{Q_{n}^{3}}(u, v)| = 2(2n-1)-1 = 4n-3$.
In the following we will prove that $Q_{n}^{3}-A_{Q_{n}^{3}}(u, v)$ is connected.

Given a decomposition of $Q_n^3$: $Q_n^3=0Q_{n-1}^{3}\bigoplus 1Q_{n-1}^{3}\bigoplus 2Q_{n-1}^{3}$ such that
$u,v$ are not in the same 3-ary $(n-1)$-subcube. Without loss of generality, suppose
$u \in V(0Q_{_{n-1}}^{3})$ and $v \in V(1Q_{_{n-1}}^{3})$.
By Lemma \ref{l4}, both $0Q_{n-1}^{3}-A_{0Q_{n-1}^{3}}(u)$ and $1Q_{n-1}^{3}-A_{1Q_{n-1}^{3}}(v)$ are connected. Let
$w$ be the common neighbor of $u, v$ in $2Q_{n-1}^{3}$.
Then $2Q_{n-1}^{3}-\{w\}$ is connected since $\kappa(Q_{n-1}^{3}) = 2n-2>1$ for $n\ge 2$.
By the definition of $Q_n^3$,
any vertex in both $V(0Q_{n-1}^{3})-A_{0Q_{n-1}^{3}}(u)$ and $V(1Q_{n-1}^{3})-A_{1Q_{n-1}^{3}}(v)$
has a pair vertex in $V(2Q_{n-1}^{3}) - \{w\}$. Thus $Q_{n}^{3}-A_{Q_{n}^{3}}(u, v)$ is connected.
Obviously, $|Q_{n}^{3}-A_{Q_{n}^{3}}(u, v)|=3^n-(4n-3)-2\ge 2$ for $n\ge 2$.
Hence, $N_{Q_{n}^{3}}(u, v)$ is an 1-extra vertex-cut for $n \geq 2$. $\hfill\rule{1mm}{2mm}$

\begin{thm}\label{t1}
$\kappa_1(Q_{n}^{3})=4n-3$ for $n\geq 2$.
\end{thm}
\emph{Proof:}
We first prove $\kappa_1(Q_{n}^{3})\ge 4n-3$. For this purpose, We only need to show that for
any $S\subset V(Q_{n}^{3})$ with $|S| \leq 4n-4$,
if there is  no isolated vertex in $V(Q_{n}^{3})-S$, then $Q_{n}^{3}-S$ is connected.

Given a decomposition of $Q_n^3$: $Q_n^3=0Q_{n-1}^{3}\bigoplus 1Q_{n-1}^{3}\bigoplus 2Q_{n-1}^{3}$.
Let $S_{i} = V(iQ_{n-1}^{3}) \cap S$ with $i = 0, 1, 2$.
Then at least one of $|S_{0}|$, $|S_{1}|$ and $|S_{2}|$ is strictly less than $2(n-1)-1$ since $S_{0}\bigcap S_{1} \bigcap S_{2} = {\emptyset}$ and $|S_{0}|+|S_{1}|+|S_{2}| = |S| \leq 4n-4<6n-6$ for $n \geq 2$. Without loss of generality, suppose $|S_{0}| \leq 2(n-1)-1$. Then $0Q_{n-1}^{3}-S_{0}$ is connected by Lemma \ref{l3}.
Let $H=1Q_{n-1}^{3}\bigoplus 2Q_{n-1}^{3}$ and $\bar{H}=H-S$.
In the following we will prove that any vertex $u$ in $V(\overline{H})$ can be 
connected to a vertex of the connected subgraph $0Q_{n-1}^{3}-S_{0}$ in $Q_{n}^{3}-S$.

Let $u_{0}$ be $u$'s pair vertex in $V(0Q_{n-1}^{3})$. If $u_{0} \not\in S_{0}$, then we are done.
So we assume that $u_{0} \in S_{0}$. Since there is no isolated vertex in $Q_{n}^{3}-S$, there exists
a vertex $v$ adjacent to $u$ in $V(H)$. According to $u_{0}$ is $v$'s pair vertex or not,
we consider the following two case.

\textbf{Case 1 } $u_{0}$ is not $v$'s pair vertex.
Then $u,v\in V(1Q_{n-1}^{3})$ or $u,v\in V(2Q_{n-1}^{3})$.
Without loss of generality, suppose $u,v\in V(1Q_{n-1}^{3})$.
Let $v_{0}$ be $v$'s pair vertex in $V(0Q_{n-1}^{3})$. If $v_{0} \not\in S_{0}$,
then we are done. So assume that $v_0 \in S_{0}$.
By Lemma \ref{l1}, $|N_{1Q_{n-1}^{3}}(u, v)|=4(n-1)-3$.
Since $|S_0|\le 2(n-1)-1<4(n-1)-3$ for $n\ge 3$, and there exists a perfect matching between
$V(1Q_{n-1}^{3})$ and $V(0Q_{n-1}^{3})$, there exists a vertex $x \in N_{1Q_{n-1}^{3}}(u, v)$ such that
$x$ and its pair vertex $\bar{x}$ in $V(0Q_{n-1}^3)$ all don't belong to $S$.
This implies that $u$ can be connected to a vertex of the connected subgraph $0Q_{n-1}^{3}-S_0$ in $Q_{n}^{3}-S$
via a path passing through $\{u, v, x,\bar{x}\}$.

\textbf{Case 2 } $u_{0}$ is $v$'s pair vertex.
Then $u\in V(1Q_{n-1}^{3})$ and $v\in V(2Q_{n-1}^{3})$,
or $v\in V(1Q_{n-1}^{3})$ and $u\in V(2Q_{n-1}^{3})$.
Without loss of generality, suppose $u\in V(1Q_{n-1}^{3})$ and $v\in V(2Q_{n-1}^{3})$.
By Lemma \ref{l1}, $|N_H(u, v)|=|N_{Q_n^3}(u, v)-\{u_0\}|=4n-3-1=4n-4$.
However, $|S_1|+|S_2|=|S |- |S_0|\le |S |- |\{u_0\}|\le 4n-4-1=4n-5$.
Thus there is a vertex $w \in N_H(u, v)$ such that $w\not\in S$. 
Then $w$ adjacent to $u$ or $v$ in $V(H)$. Without loss of generality, 
suppose $w$ is adjacent to $v$. Then $w\in V(2Q_{n-1}^{3})$.
Let $w_{0}$ be $w$'s pair vertex in $V(0Q_{n-1}^{3})$. If $w_{0} \not\in S_{0}$, then we are done.
So we assume that $w_{0} \in S_{0}$.
By Lemma \ref{l1}, $|N_{2Q_{n-1}^{3}}(v, w)|=4(n-1)-3$. Since
$|S_0|\le 2(n-1)-1<4(n-1)-3$ for $n\ge 3$, and there exists a perfect matching between
$V(2Q_{n-1}^{3})$ and $V(0Q_{n-1}^{3})$, there
exists a vertex $x$ in $N_{2Q_{n-1}^{3}}(v,w)$ such that $x$ and its pair vertex $\bar{x}$ in $V(0Q_{n-1}^3)$
all don't belong to $S$.
This implies that $u$ can be connected to a vertex of the connected subgraph $0Q_{n-1}^{3}-S_0$ in
$Q_{n}^{3}-S$ via a path passing through $\{u, v, w, x,\bar{x}\}$.

So any vertex in $V(\bar{H})$ can be connected to a vertex of the connected subgraph $0Q_{n-1}^{3}-S_{0}$ in $Q_{n}^{3}-S$. Therefore, $Q_{n}^{3}-S$ is connected.
Thus $\kappa_1(Q_{n}^{3})\ge 4n-3$ for $n \geq 2$.
However, by Lemma \ref{l5}, $\kappa_1(Q_{n}^{3})\le 4n-3$ for $n \geq 2$.

Hence, $\kappa_1(Q_{n}^{3})=4n-3$ for $n \geq 2$.

$\hfill\rule{1mm}{2mm}$

% \subsection{A simulation example}

\section{2-extra connectivity}
\begin{lem}\label{l6}
$\kappa_2(Q_{n}^{3})\le 6n-7$ for $n \geq 3$.
\end{lem}
\emph{Proof:}
Suppose $P = (u, v, w)$ is a path of length two in $Q_{n}^{3}$ and $(u, w) \not\in E(Q_{n}^{3})$.
Then $Q_{n}^{3}-N_{Q_{n}^{3}}(P)$ is disconnected.
By Lemma \ref{l1}, $u$ and $v$ (resp. $v$ and $w$) have exactly one common neighbor $x$ (resp. $y$), and $u$ and $w$ have exactly two common neighbors $v,z$. By the definition of $Q_n^3$, $z$ is different from
$x$ and $y$ (see Figure 2 (a) for illustration).
Thus, $|N_{Q_{n}^{3}}(P)| = 6n-7$.
In the following we will prove that $Q_{n}^{3}-A_{Q_{n}^{3}}(u, v, w)$ is connected.

\begin{figure}%[http]
\centering
\scalebox{1.0}{\includegraphics{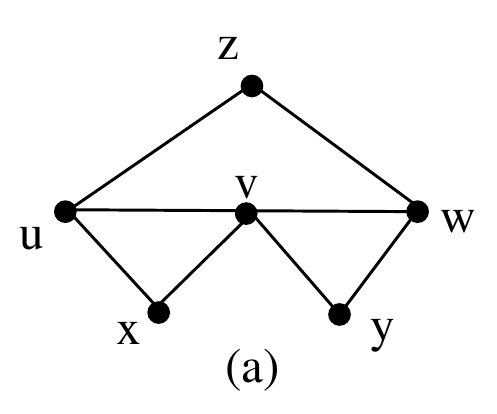}}
\scalebox{0.8}{\includegraphics{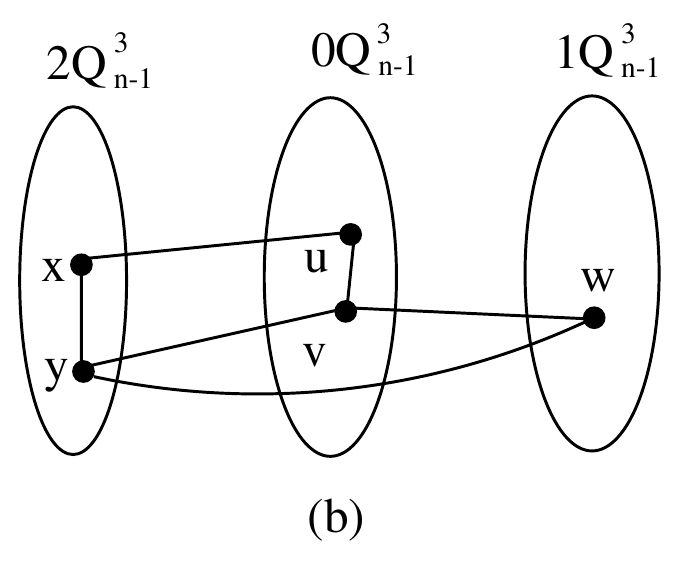}}
\caption{Illustration for Lemma \ref{l6}}
%\label{fig:anna}
\end{figure}

Given a decomposition of $Q_n^3$: $Q_n^3=0Q_{n-1}^{3}\bigoplus 1Q_{n-1}^{3}\bigoplus 2Q_{n-1}^{3}$ such that
$(u, v)\in E(0Q_{n-1}^{3})$, and $w$ is $v$'s pair vertex in $V(1Q_{n-1}^{3})$.
By Lemma \ref{l1}, $u$ has a pair vertex $x$ in $V(2Q_{n-1}^{3})$,
and $v$ and $w$ have a common neighbor $y$ in $V(2Q_{n-1}^{3})$.
Clearly, $2Q_{n-1}^3- \{x,y\}$ is connected for $n\ge 3$. By the definition of $Q_n^3$,
any vertex in both $V(0Q_{n-1}^3)- A_{0Q_{n-1}^3}(u,v)$ and $V(1Q_{n-1}^3)- A_{1Q_{n-1}^3}(w)$
has a neighbor in $V(2Q_{n-1}^3)- \{x,y\}$ (see Figure 2 (b) for illustration). Thus $Q_{n}^{3}-A_{Q_{n}^{3}}(u, v, w)$ is connected (see Fig. 3).
Obviously, $|Q_{n}^{3}-A_{Q_{n}^{3}}(u,v,w)|=3^n-(6n-7)-3>3$ for $n\ge 3$.
So $N_{Q_{n}^{3}}(u,v,w)$ is an 2-extra vertex-cut for $n \geq 3$.

Hence, $\kappa_2(Q_{n}^{3})\le 6n-7$ for $n \geq 3$.
$\hfill\rule{1mm}{2mm}$

\begin{thm}\label{t2}
$\kappa_2(Q_{n}^{3})\ge 6n-7$ for $n \geq 3$.
\end{thm}
\emph{Proof:}
For this purpose, we only need to prove that for any $S\subset V(Q_{n}^{3})$ with $|S| \leq 6n-8$, if
there is no isolated vertex or isolated edge in $Q_{n}^{3}-S$, then $Q_{n}^{3}-S$ is connected.

Given a decomposition of $Q_n^3$: $Q_n^3=0Q_{n-1}^{3}\bigoplus 1Q_{n-1}^{3}\bigoplus 2Q_{n-1}^{3}$.
Let $S_{i} = V(iQ_{n-1}^{3}) \cap S$ with $i = 0, 1, 2$.
Then at least one of $|S_{0}|$, $|S_{1}|$ and $|S_{2}|$ is not more than $2(n-1)-1$ since $S_{0}\bigcap S_1\bigcap S_{2} = {\emptyset}$ and $|S_{0}|+|S_{1}|+|S_{2}| = |S| \leq 6n-8$. Without loss of generality, suppose that $|S_{0}| \leq 2(n-1)-1$. Then $0Q_{n-1}^{3}-S_{0}$ is connected by Lemma \ref{l3}.
Let $H=1Q_{n-1}^{3}\bigoplus 2Q_{n-1}^{3}$ and $\bar{H}=H-S$.
In the following we will prove that any vertex $u$ in $V(\bar{H})$ can be connected to a vertex of the connected subgraph $0Q_{n-1}^{3}-S_{0}$ in $Q_{n}^{3}-S$.

Let $u_{0}$ be $u$'s pair vertex in $V(0Q_{n-1}^{3})$. If $u_{0} \not\in S_{0}$, then we are done. So assume that $u_{0} \in S_{0}$. Since there is no isolated vertex in $Q_{n}^{3}-S$, there exists a vertex $v$ adjacent to $u$ in $V(\bar{H})$. According to $u_{0}$ is $v$'s pair vertex or not, we consider the following two cases.

\textbf{Case 1 } $u_{0}$ is not $v$'s pair vertex.
Then $u,v\in V(1Q_{n-1}^{3})$ or $u,v\in V(2Q_{n-1}^{3})$.
Without loss of generality, suppose $u,v\in V(1Q_{n-1}^{3})$.
Let $v_{0}$ be $v$'s pair vertex in $V(0Q_{n-1}^{3})$. If $v_{0} \not\in S_{0}$,
then we are done. So assume that $v_0 \in S_{0}$.
By Lemma \ref{l1}, $|N_{1Q_{n-1}^{3}}(u, v)|=4(n-1)-3$.
Since $|S_0|\le 2(n-1)-1<4(n-1)-3$ for $n\ge 3$, and there exists a perfect matching between
$V(1Q_{n-1}^{3})$ and $V(0Q_{n-1}^{3})$, there exists a vertex $x \in N_{1Q_{n-1}^{3}}(u, v)$ such that
$x$ and its pair vertex $\bar{x}$ in $V(0Q_{n-1}^3)$ all don't belong to $S$.
This implies that $u$ can be connected to a vertex of the connected subgraph $0Q_{n-1}^{3}-S_0$ in $Q_{n}^{3}-S$
via a path passing through $\{u, v, x,\bar{x}\}$.

\textbf{Case 2 } $u_{0}$ is $v$'s pair vertex. Then $u\in V(1Q_{n-1}^{3})$ and $v\in V(2Q_{n-1}^{3})$,
or $v\in V(1Q_{n-1}^{3})$ and $u\in V(2Q_{n-1}^{3})$.
Without loss of generality, suppose $u\in V(1Q_{n-1}^{3})$ and $v\in V(2Q_{n-1}^{3})$.
Since there is no isolated edge in $Q_{n}^{3}-S$, there exists a vertex $w$ adjacent to $u$ or $v$ in $V(H)$. Without loss of generality, suppose $w$ is adjacent to $v$. Then $w\in V(2Q_{n-1}^{3})$
Let $w_{0}$ be $w$'s pair vertex in $V(0Q_{n-1}^{3})$. If $w_{0} \not\in S_{0}$, then we are done.
So we assume that $w_{0} \in S_{0}$.
By Lemma \ref{l1}, $|N_{2Q_{n-1}^{3}}(v, w)|=4(n-1)-3$. Since
$|S_0|\le 2(n-1)-1<4(n-1)-3$ for $n\ge 3$, and there exists a perfect matching between
$V(2Q_{n-1}^{3})$ and $V(0Q_{n-1}^{3})$, there
exists a vertex $x$ in $N_{2Q_{n-1}^{3}}(v,w)$ such that $x$ and its pair vertex $\bar{x}$ in $V(0Q_{n-1}^3)$
all don't belong to $S$.
This implies that $u$ can be connected to a vertex of the connected subgraph $0Q_{n-1}^{3}-S_0$ in 
$Q_{n}^{3}-S$ via a path passing through $\{u, v, w, x,\bar{x}\}$.

So any vertex in $V(\bar{H})$ can be connected to a vertex of the connected subgraph $0Q_{n-1}^{3}-S_{0}$ in $Q_{n}^{3}-S$. Therefore, $Q_{n}^{3}-S$ is connected. 

Hence, $\kappa_2(Q_{n}^{3})\ge 6n-7$ for $n \geq 3$.

$\hfill\rule{1mm}{2mm}$

By Lemma \ref{l6} and Theorem \ref{t2}, we obtain the following corollary.
\begin{cor}
$\kappa_2(Q_{n}^{3})=6n-7$ for $n \geq 3$.
\end{cor}

\section{Conclusions and future research}
In this paper, we have obtained, for the first time, the 1-extra connectivity (the super connectivity) and the 2-extra connectivity of 3-ary $n$-cubes.
The main results of this paper are as follows: $$\kappa_1(Q_{n}^{3}) = 4n-3 \mbox{ for } n\ge 2; \ \ \ \  \kappa_2(Q_{n}^{3}) = 6n-7 \mbox{ for }n\ge 3.$$
The tools we have used to determine the 1-extra connectivity and the 2-extra connectivity of 3-ary $n$-cubes is Lee distance.  Using Lee distance we have showed that 3-ary $n$-cubes can be decomposed into 3 vertex disjoint $(n-1)$-subcubes along any dimension. We believe this property will be useful in the study of other properties of 3-ary $n$-cubes, for example,   the diagnosability and conditional diagnosability of 3-ary $n$-cubes both  under the PMC model and Comparison model. In the future, we will explore the conditional diagnosability of $k$-ary $n$-cubes both under the PMC model and Comparison model.

\section*{References}
\bibliographystyle{elsarticle-num}
\bibliography{parallelcomputing}
%% Authors are advised to submit their bibtex database files. They are
%% requested to list a bibtex style file in the manuscript if they do
%% not want to use elsarticle-harv.bts.

%% References without bibTeX database:

%\begin{thebibliography}{99}
%% \bibitem must have one of the following forms:
%%   \bibitem[Jones et al.(1990)]{key}...
%%   \bibitem[Jones et al.(1990)Jones, Baker, and Williams]{key}...
%%   \bibitem[Jones et al., 1990]{key}...
%%   \bibitem[\protect\citeauthoryear{Jones, Baker, and Williams}{Jones
%%       et al.}{1990}]{key}...
%%   \bibitem[\protect\citeauthoryear{Jones et al.}{1990}]{key}...
%%   \bibitem[\protect\astroncite{Jones et al.}{1990}]{key}...
%%   \bibitem[\protect\citename{Jones et al., }1990]{key}...
%%   \harvarditem[Jones et al.]{Jones, Baker, and Williams}{1990}{key}...
%%

% \bibitem[ ()]{}

%\end{thebibliography}

\end{document}